# The effect of the thermal reduction on the kinetics of low temperature $^4$He sorption and the structural characteristics of graphene oxide.

A.V. Dolbin[1], M.V. Khlistuck[1], V.B. Esel'son[1], V.G. Gavrilko[1], N.A. Vinnikov[1], R.M. Basnukaeva[1], A.I. Prokhvatilov[1], I.V. Legchenkova[1], V.V. Meleshko[1],

W.K. Maser[2], A.M. Benito[2]

[1] B. Verkin Institute for Low Temperature Physics and Engineering of the National Academy of Sciences of Ukraine 47, Lenin Ave., 61103 Kharkov, Ukraine

[2] Instituto de Carboquímica, 4, ICB-CSIC Miguel Luesma Castán, 4 E-50018 Zaragoza, Spain

Electronic address: dolbin@ilt.kharkov.ua

**Abstract.**

The kinetics of the sorption and the subsequent desorption of $^4$He by the starting graphite oxide (GtO) and the thermally reduced graphene oxide samples (TRGO, $T_{reduction}$ = 200, 300, 500, 700 and 900$^0$ C) have been investigated in the temperature interval 1.5 – 20 K. The effect of the annealing temperature on the structural characteristics of the samples was examined by the X-ray diffraction (XRD) technique. On lowering the temperature from 20 K to 11-12 K, the time of $^4$He sorption increased for all the samples, which is typically observed under the condition of thermally activated diffusion. Below 5 K the characteristic times of $^4$He sorption by the GtO and TRGO-200 samples were only weakly dependent on temperature, suggesting the dominance of the tunnel mechanism. In the same region (T<5 K) the characteristic times of the TRGOs reduced at higher temperatures (300, 500, 700 and 900$^0$ C) were growing with lowering temperature, presumably due to the defects generated in the carbon planes on removing the oxygen functional groups (oFGs). The estimates of the activation energy ($E_a$) of $^4$He diffusion show that in the TRGO-200 sample the $E_a$ -value is 2.9 times lower as compared to the parent GtO, which is accounted for by GtO exfoliation due to evaporation of the water intercalated in the interlayer space of carbon. The nonmonotonic dependences $E_a(T)$ for the GtO samples treated above 200$^o$ C are determined by a competition between two processes – the recovery of the graphite carbon structure, which increases the activation energy, and the generation of defects, which decreases the activation energy by opening additional surface areas and ways for sorption. The dependence of the activation energy on $T_{treatment}$ correlates well with the contents of the crystalline phase in GtO varying with a rise of the annealing temperature.

## 1. Introduction.

Graphene is a two-dimensional system of carbon atoms in which three electrons of each C atom form strong hybridized sp$^2$ bonds in the plane that build a honeycomb structure; the fourth π - electron is smeared below and above the carbon layer. The π - electrons are very important as they form π- electron bonds and largely determine the properties of multilayered graphenes and graphene oxides [1,2]. Owing to its two-dimensional structure and electron hybridization, graphene possesses unique properties, such as ballistic electrical conductivity and high intrinsic carrier mobility (200 000 cm$^2$ v$^{-1}$ s$^{-1}$) [3,4], high thermal conductivity (~ 5000 W m$^{-1}$ K$^{-1}$) [6], extraordinarily high mechanical characteristics (Young's modulus of ~1.0 TPa [15]) and the quantum Hall effect [7]. Besides, graphene has a large specific surface area (2630 m$^2$ /g) and can be used as a highly efficient sorbent. Graphene can serve as a basis for other allotropic carbon modifications: it can be rolled up to form 0D fullerenes, or twisted to produce 1D carbon nanotubes, or packed as 3D graphite [2]. At present there exist a large variety of mechanical [8], physical [9] and chemical [10] methods for obtaining graphene. The modified Hummers method [11, 12] is most advantageous for cost-effective large-scale production of graphene oxide (GO) through oxidation – induced exfoliation of graphite. However, the resulting graphene sheets are greatly overloaded at both sides with oxygen inherent in the added hydroxyl, epoxy, carboxyl and other oxygen functional groups (oFGs) [13, 14]. The cFGs contents in GO can be decreased considerably through chemical and thermal treatment [15, 16]. GO holds much promise as a stock for large-scale production of grephene–based materials and their widespread applications in advanced industrial technologies, such as photovoltaic cells, capacitors, sensors and transparent conductive electrodes [2, 17-19]. Owing to their advantageous characteristics, such as low density, chemical stability, diversity of structural forms, easy-to-modify porous structure, surface susceptible to a variety of treatment techniques and relatively simple technologies of industrial-scale production, graphene – based materials hold considerable potential for development of gas storage technologies, adsorption/desorption of impurity particles among them.

Previously we investigated the sorption and the subsequent desorption of $^4$He, $H_2$, Ne, $N_2$, $CH_4$ and Kr by glucose- and hydrazine-reduced GO [20]. It was found that in the temperature interval 2-290 K, the temperature dependences of the diffusion coefficients of light impurities (hydrogen and helium) were controlled by a competition between the thermally activated and tunnel mechanisms of diffusion. The contribution of the tunnel process is dominant at low temperatures, which makes the diffusion coefficients practically independent of temperature. However, the tunnel effects were less pronounced for heavier impurities ($N_2$, $CH_4$ and Kr). Tunnel diffusion of light impurities at low temperatures was also observed in other carbon nanostructures, specifically in carbon nanotubes [21, 22] and fullerite $C_{60}$ [23-25]. This led us to assume that the kinetics of gas sorption by carbon nanostructures was significantly influenced by the potential relief of their surfaces. Furthermore, it was found that oFGs also had a significant effect on the sorption properties of graphene oxide [20]: the sorption capacity of GO increased three- to six-fold after removing oFGs through hydrazine-reduction. The result suggests that the removal of oFGs via hydrazine reduction unblocks the interlayer space in GO and allows the gas impurities to penetrate inside through the defects on the graphene surface.

The condition of the carbon surface of GO depends strongly on the type, quantity and distribution of oFGs, the degree and the method of their removal (GO reduction) as well as on the quantity and the character of defects generated by the oFGs removal. Investigation of the reduction effect on the sorption characteristics of GO will help us to trace the structural evolution during GO reduction and to find the ways of modifying the RGO (reduced GO) properties. There exist two basic approaches – chemical and thermal- that can ensure a high degree of reduction.

At present thermally reduced graphene (TRGO) is prepared from graphite oxide (GtO) [27, 28] which is in turn obtained from graphite using various chemical oxidizing agents [11, 29, 30]. The thermal reduction of graphene attracts interest because, on the one hand, it offers the advantage of cost-effective large-scale production of graphene and, on the other hand, the thermally reduced graphene is free from the chemical residuals unremovable in GO powders and films [28, 31]. Note that on chemical reduction some of the GO sheets in solutions contain chemically strong bases (for example, hydrazine) and are therefore unsuitable for biological and medical applications [22, 32]. Thermal reduction is, however, a complex process involving a thermally activated multistep removal of intercalated $H_2O$ molecules and oxide functional carboxyl (COOH), hydroxyl
 (C-OH) and epoxy (C=O) groups, including interlayer epoxy (C-O-C) and single-bonded (C=O) ones at the carbon plane surface. The functional groups are located both on the graphene surface and at the plane edges. The reduction process is aimed at removing oFGs and obtaining a carbon structure with the desired properties. It should be noted that oxidizing treatment generates numerous defects and surface ruptures in GO. In the course of GO reduction the number of imperfections increases since the removal of oFGs often strips the carbon atoms off the graphene plane. Besides, the $C-O_x$ type $sp^3$ – bonded oxide groups can be present inside the defects of GO and GtO. The interlayer spacings are considerably larger in GO than in GtO due to the intercalated $H_2O$ molecules and the presence of various oxide groups (9-12 A depending on the method of GO preparation and the amount of the intercalated water against 7 A for GtO [11]).

Previously the sorbed quantities of $^4$He, $H_2$, Ne, $N_2$ and Kr were investigated as a function of the thermal reduction temperature of GO [26, 34]. The largest quantities of these impurities were sorbed by the samples reduced at 300 and 900º C. It was assumed that the sorption capacity of the samples reduced at 300º C grew higher because of the disordering of the layered GO structure inflicted by water de-intercalation. The higher sorption capacity of the sample heated to 900º C was attributed to the generation of numerous defects at the carbon surfaces on removing the oFGs, which allowed the gas impurities to penetrate into the space between the folds and sheets of the graphene structure [26].

Here we have investigated the effect of the reduction temperature upon the kinetics of the low temperature $^4$He sorption and the structural characteristics of graphene oxide.

## 2. Samples and their characterization by the X-ray diffraction method.

The starting graphite oxide (GtO) was prepared from graphite powder (Sigma-Aldrich) using the modified Hummers method and vigorous oxidizing agents ($NaNO_3$, $H_2SO_4$ and $KMnO_4$) [11, 35]. The resulting product was then thermally treated under Ar atmosphere to prepare thermally reduced graphene oxide (TRGO). Five samples were obtained by heating the starting GtO material at 200, 300, 500, 700 and 900ºC, respectively [26]. The GtO and reduced TRGO samples were powders with an average grain size about 10 μm, the mass of each sample was ~40 mg.

The effect of the annealing temperatures upon the structural characteristics of GtO, was investigated by the X-ray diffraction (XRD) method in Cu-K$_\alpha$ radiation using a DRON-3 diffractometer. The obtained XRD patterns (Fig.1) are similar qualitatively to those in [26] but for a few distinctions (a detailed analysis and a comparison of the XRD patterns with other data were rather difficult because of the lack of quantitative data processing in [26]).

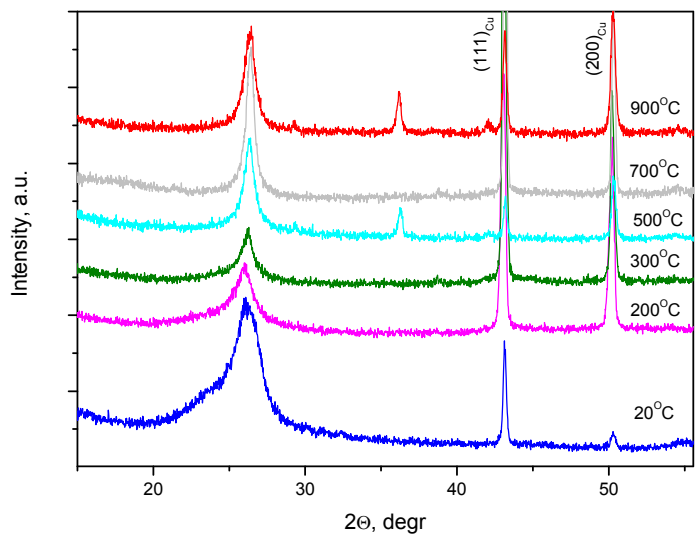

Fig. 1. The XRD patterns of GtO powder at different annealing temperatures (indicated at the end of each XRD pattern). The XRD patterns taken in Cu-K$_\alpha$ radiation contain (111) and (200) reflections from a copper substrate.

It is seen that oxidation causes serious damage to the graphite structure (Fig. 1, T=20$^o$ C). The intensive sharp peaks (002) in the region 2$\Theta$ = 26$^o$ produced by the basal planes of the hexagonal close-packed lattice of pure graphite disappear. Instead, the starting GtO exhibits a broadened asymmetric medium-intensity line. The considerable asymmetry of the reflection most likely betokens the presence of another carbon phase. The analysis of the reflections shows that the obtained diffractogram can be described quite adequately by a sum of the intensities of the two lines (Fig.2) corresponding to diffraction from two nanocarbon phases having significantly different degrees of crystallinity. The small-angle reflection has the structural parameters that are typical for amorphous states. The angular smearing of the line runs to several degrees, which points to severe distortions of the structure in the short-range order region of the basal planes causing numerous static interlayer displacements $\Delta d_{00\,l}$ in this phase as compared to the mean interlayer spacing in pure graphite.

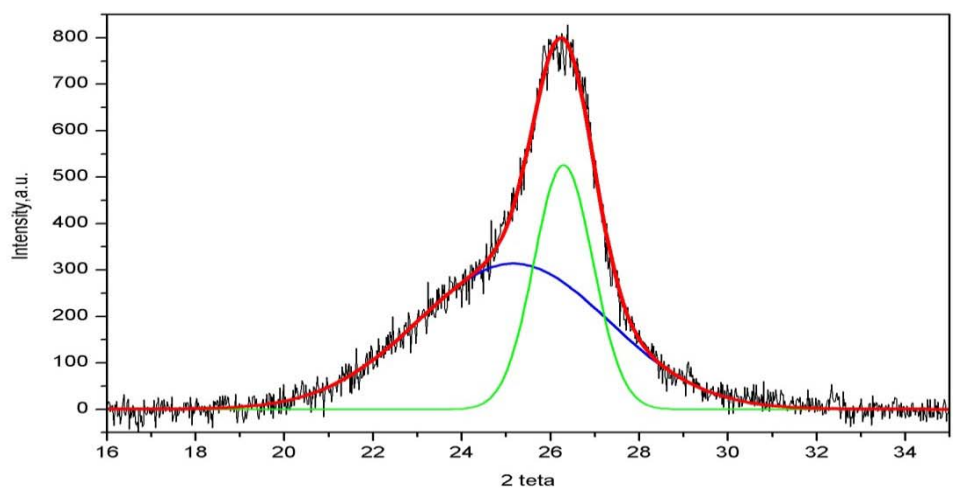

Fig. 2. The separated intensity contributions of XRD from the (002) planes of the crystalline GtO (green) and 'amorphous' GO (blue) phases at room temperature (20°).

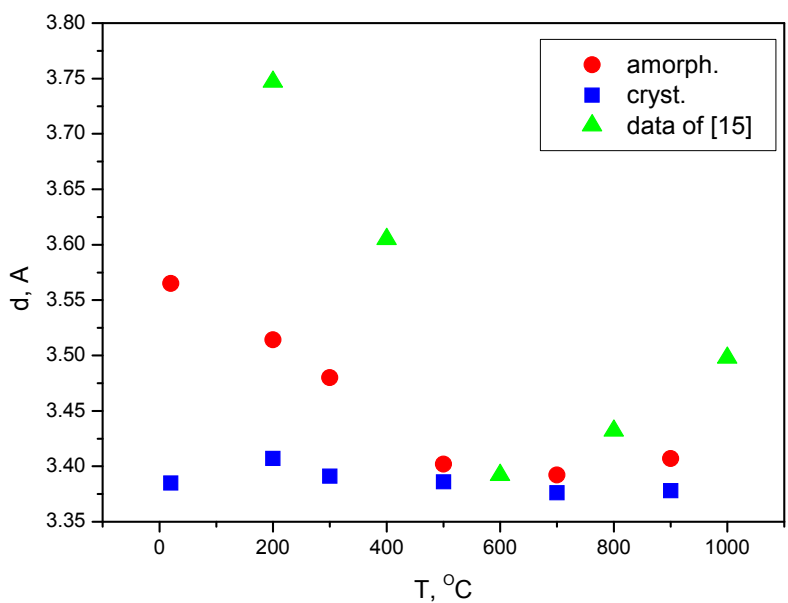

Fig. 3. The averaged interplane spacings $d_{002}$ in the crystalline (■) and amorphous (●) phases of the starting GtO and TRGO at different annealing temperatures. Comparison with the data of [15] (▲).

The second phase of GtO has a higher degree of crystallinity. Like pure graphite, it has a hexagonal close-packed (HCP) structure. But the diffraction reflections from this phase are broader than those from graphite single crystals, being however almost three times narrower in comparison with the disordered latent-crystalline (amorphous) modification (Fig.2). These facts suggest that the crystalline phase of the starting graphite is not perfect structurally.

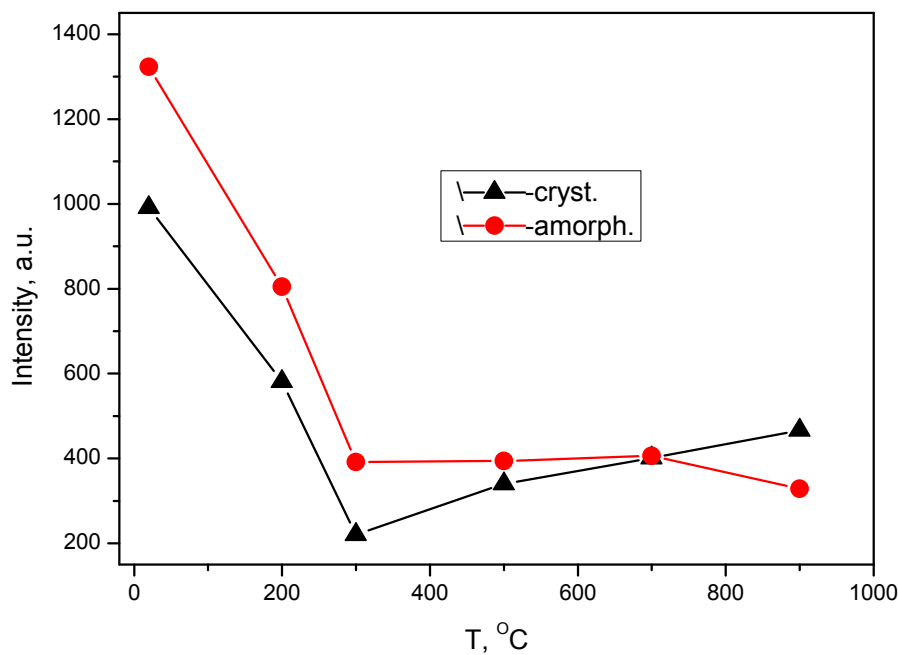

Fig.4. The effect of annealing temperatures on the integral intensity of XRD reflections (002) from the crystalline (▲) and amorphous (•) phases of GtO and GO.

The XRD results on the effect of high temperature annealing upon the structural characteristics of GtO and TRGO are shown in Figs. 1 and 3-7 . The general outline in Fig.1 suggests that a rise of the annealing temperature causes appreciable transformations in the structure of the samples, which results in reduction of graphite. The diffraction reflections from the basal planes (002) of both phases shift towards the larger angle region and their width and asymmetry decrease. It was surprising to observe a distinct narrow reflection (Fig.1) at $T_{anneal}$ = 500° C in the angular region 2Θ = 36°, which had nothing to do with either crystalline or amorphous graphite. Among the known carbon forms, only cubic and hexagonal diamonds can produce reflections at these angles. It is then reasonable to assume that the annealing of GtO at T ≥ 500° C causes $sp^3$ hybridization of carbon in individual regions of the GtO (graphene) crystallites. However, this assumption did not agree with the subsequent results: the line in question was absent at $T_{anneak}$ = 700° C but reappeared at T= 900° C (fig.1). Thus, the nature of the coherent diffraction at 36° remains uncertain and calls for further investigation by XRD and other methods, including spectroscopy.

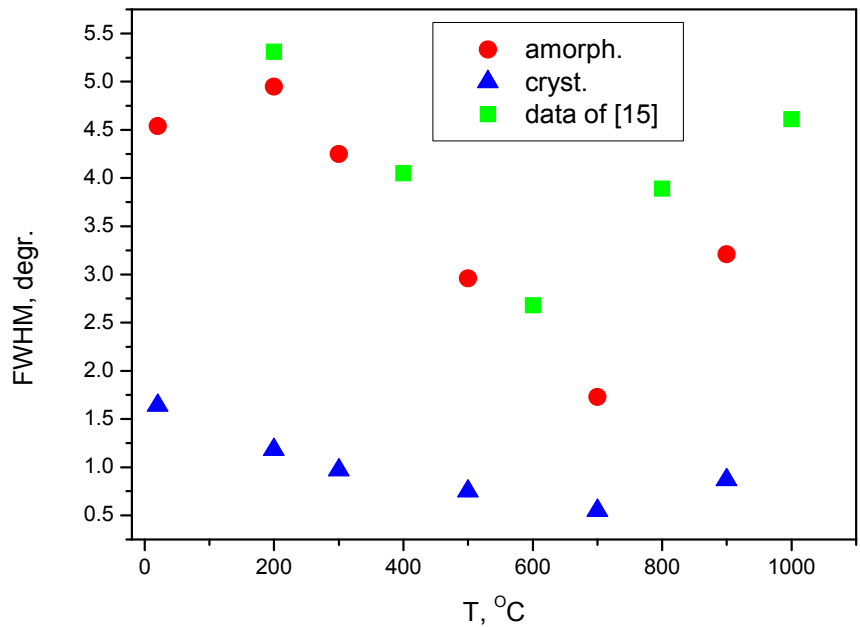

Fig. 5. Variations of the maxima half-widths of scattering diffraction from the (002) planes of the crystalline (▲) and amorphous (•) phases of graphite oxide at different annealing temperatures. Green symbols (■) are XRD data of [15].

The quantitative changes in the structural parameters (interplanar spacing $d_{002}$, integral intensity and diffraction line half-width) of both phases are illustrated in Figs. 3 – 6. As the annealing temperature rises, the parameter magnitudes decrease for both the phases, then approach each other (Figs. 3, 6) and at T>500º C reach the values typical for pure graphite. The structural characteristics of the latent crystalline (amorphous) phase are particularly sensitive to the annealing temperature. Note that heating to T=900º C does not ensure complete reduction of graphite. Even at this high temperature the samples still contain an appreciable quantity of the amorphous phase formed at the starting stage of graphite oxidation. Moreover, as is evident from our results and the data of [15], the temperature dependence of the structural parameters (interplanar spacing $d_{200}$ (Fig.3) and the XRD reflection half-widths (Fig.)) suffers inversion at T>500º C presumably due to the thermal defects formed and accumulated in the carbon sublattice. However, this process has only a minor effect on the intensity of diffraction from the crystalline phase and its quantity increases monotonically, as against the amorphous phase, in the whole interval of annealing temperatures (Fig.7).

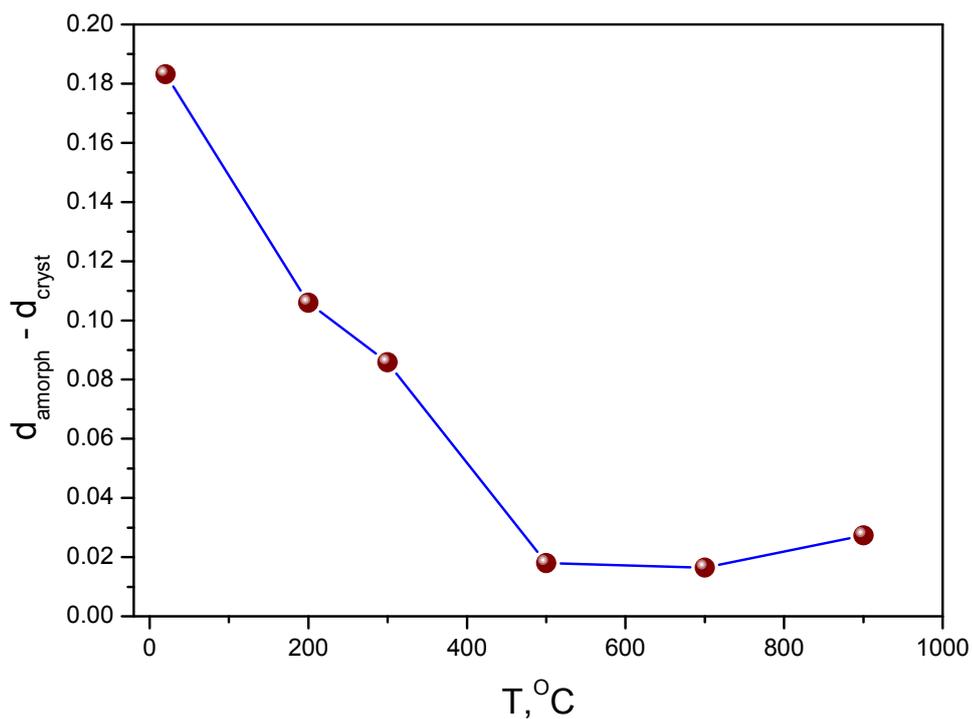

Fig. 6. The effect of the GtO annealing temperatures on the difference between the interplanar spacings $d_{002}$ in the amorphous and crystalline phases.

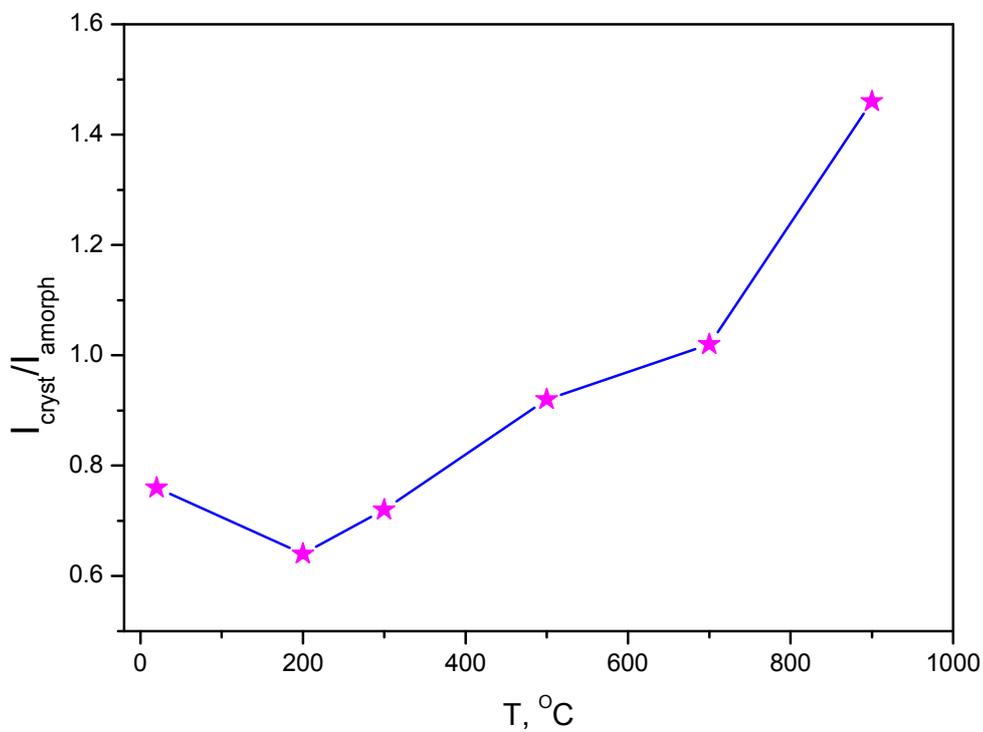

Fig. 7. The growth of the crystalline GtO phase with increasing annealing temperature characterized by a ratio between integral intensities $I_{cryst}/I_{amorph}$.

It is known that thermal treatment is one of the efficient methods of obtaining reduced graphene oxide. We believe that the amorphous phase observed at high annealing temperatures contains a great quantity of multilayered graphene, or rather TRGO, flakes. The assumption is consistent with our results (see above) on the structural parameters of the latent crystalline (amorphous) phase which in turn agree well qualitatively and quantitatively with literature XRD data [15, 16, 26] (Figs. 3-8). At T>900° C the carbon phase is almost completely reduced, the $O_2$ contents being below 3% (see Fig.8).

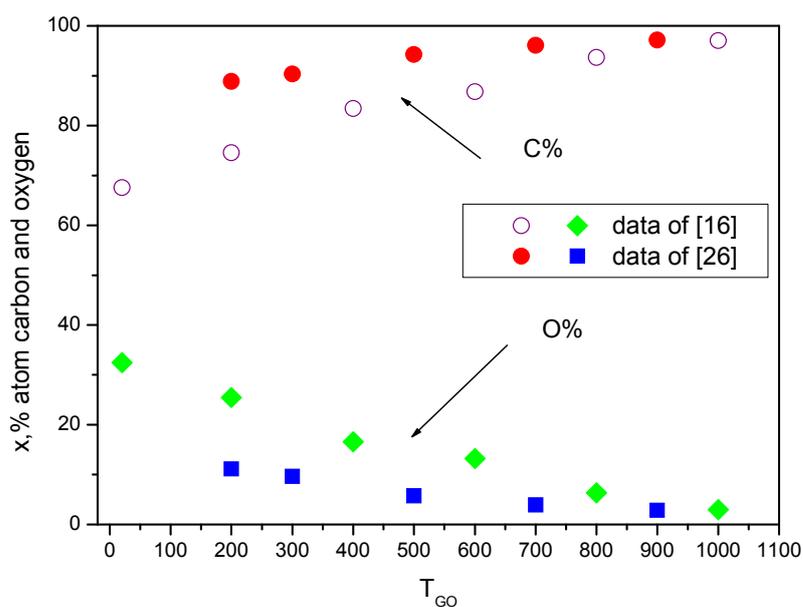

Fig. 8. The effect of annealing temperatures on the elemental GO composition according to the XPS data of [16] (○,♦) and [26] (■, ●).

Besides, our XRD patterns exhibit a weak feature in the region $2\Theta = 10°$, which can be interpreted as X-ray diffraction by the quasi-two-dimensional carbon graphene phase. This value (10°) corresponds to (001) of graphene oxide with the intercalated water, and in our case, perhaps is due to the presence of trazes of non reduced graphene oxide. Unfortunately, the growing background of the initial X-ray beam in this region of diffraction angles makes it difficult to characterize this phase quantitatively.

**3. Experimental desorption technique.**

The kinetics of sorption and desorption of $^4$He gas by the starting graphite oxide (GtO) and thermally reduced graphene oxide (TRGO) was investigated through measuring the time – pressure dependence of the gas contacting the sample in a closed vessel. The experimental technique and equipment are detailed in [23, 36, 37]. Prior to investigation each sample was placed into a measuring cell and kept in vacuum no lower than $10^{-4}$ Torr for a week to remove gas impurities. The cell was washed three times with dry nitrogen for a more efficient removal of water vapor. The impurity contents in $^4$He gas were no more than 0.002%. The sample was saturated with the $^4$He impurity under a pressure of ~1 Torr. The lowest temperature of the experiment (up to 1.5 K) was dictated by the setup design. In the whole range of the experimental temperatures the $^4$He pressure in the measuring cell was maintained 2.5-3 times lower than the saturation vapor pressure for this impurity. The temperature conditions prevented condensation of $^4$He vapor on the sample surface and the cell walls. As the impurity was adsorbed, new portions of $^4$He gas were added. The gas supply was cut off on reaching the equilibrium pressure ($10^{-2}$ Torr) in the cell.

At each step of saturation and desorption measurement the sample was kept at a pre-assigned invariant temperature. The pressure variations of the gas in the closed vessel with the sample were measured during saturation/desorption using Baratron MKS capacitance pressure transducers, the error being no more than 0.05%. When the process of sorption was completed, the cell was hermetically sealed, and the pressure variations were registered in the process of the impurity desorption from the powder on its stepwise heating. The gas impurity released on heating was taken in portions to an evacuated calibrated vessel. The gas extraction from the samples was continued until the gas pressure over the samples decreased to $10^{-2}$ Torr. Then the process of desorption was repeated at the next temperature point.

## 4. Results and discussion.

The kinetics of the helium sorption and the subsequent desorption by the starting graphite oxide and graphene oxide thermally reduced at different reduction temperatures has been investigated in the temperature interval 1.5 – 20 K. The obtained time dependences of the $^4$He pressure variations in the cell with the sample in the course of sorption/desorption are described well by the exponential one–parameter function ($\tau$) (see Fig. 9).

$$\Delta P = A(1 - \exp(-t/\tau)) \qquad (1),$$

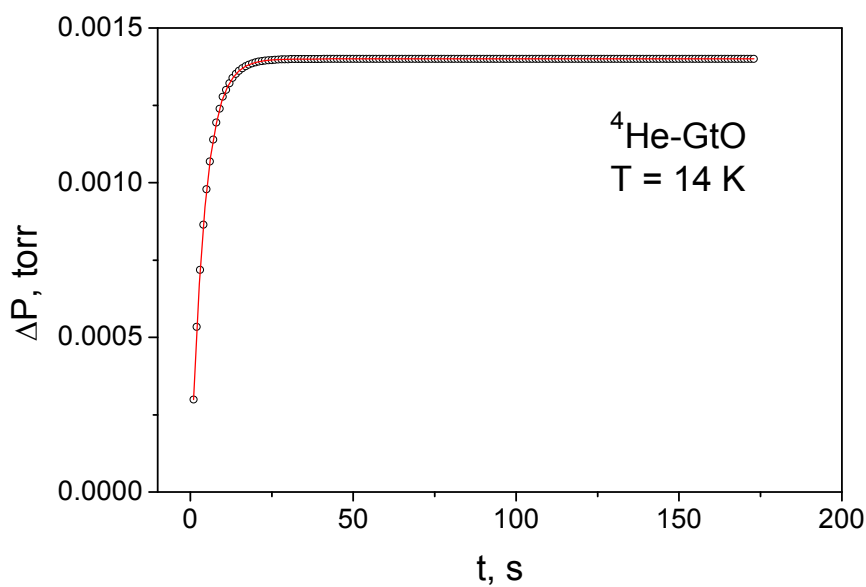

Fig. 9. The pressure variations in the course of $^4$He desorption from the GtO sample (symbols) and their description with the exponential function, Eq. (1) (line). The results measured at T=14 K are presented as an example.

At a constant temperature the characteristic times of sorption/desorption measured on the same sample coincide within the experimental error. The temperature dependences of the characteristic times of $^4$He sorption/desorption are illustrated in Fig. 10 for GtO (Fig. 10a) and TRGOs ( Figs. 10a, b). Note that in the whole interval of the temperatures used in the experiment the measurement error induced by the proper time taken by the gas phase to reach the thermal equilibrium in the measuring gas system (thermalization time) was at least an order of magnitude lower for all the samples than the measured characteristic times.

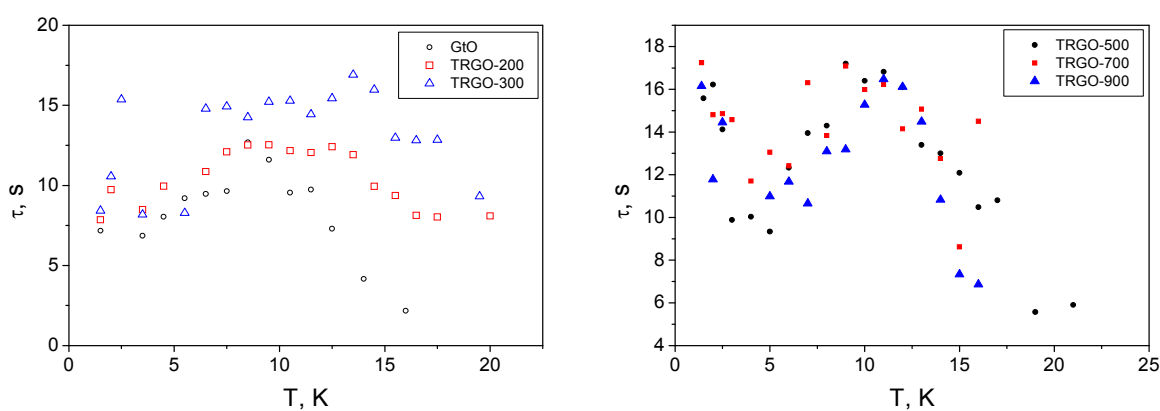

Fig. 10. The temperature dependences of the characteristic times of $^4$He sorption by a) GtO, TRGO-200, TRGO-300 and b) TRGO-500, TRGO-700, TRGO-900 samples.

The times of $^4$He sorption increased for all samples when the temperature was lowered from 20 K to about 11-12 K (see Figs. 10 a,b). This behavior indicates that in this temperature interval the sorption is mainly controlled by thermally activated diffusion of $^4$He atoms. On a further drop of temperature the sorption times started to decrease and at T<5 K the characteristic times of $^4$He sorption by the GtO and TRGO-200 samples were only little dependent on temperature (Fig. 10a ). The observed features suggest that below 5 K the sorption/desorption rate is determined by the dominant process of $^4$He atom tunneling between the carbon planes of GO. The nonmonotonic behavior of the temperature dependences of the characteristic times of the $^4$He sorption by the GtO and TRGO-200 samples is most likely accounted for by a competition between thermally activated diffusion dominant at T>12 K and the tunneling process prevailing at low temperatures. Similar effects were also observed while investigating the gas sorption by fullerite $C_{60}$, single-walled carbon nanotubes [21, 22] and chemically reduced graphene [20]. The tendency for a growth of the characteristic times of $^4$He sorption with the thermal treatment temperature observed for the TFGO-200 and TRGO-300 samples, as against the starting GtO (Fig. 10 a), is most likely related to the structural and morphological changes occurring as the water intercalated in the interlayer spacings of carbon is evaporated in the course of thermal treatment [26].

Below 5 K the characteristic times of GO samples thermally reduced at higher temperatures (TRGO-300, TRGO-500, TRGO-700 and TRGO-900) increased with lowering temperature (Figs. 10 a, b). This can be due to the growing quantity of defects in the carbon planes on thermally stimulated removal of oxygen functional groups (oFGs) [26]. They form additional potential barriers impeding diffusion at the defect locations and diminishing the probability of tunneling.

The obtained τ- values were used to estimate the coefficients of helium diffusion into GtO and thermally reduced GO:

$$D \approx \frac{\overline{\ell}^2}{4 \cdot \tau} \quad , (2)$$

where $\overline{\ell}$ is the mean grain size of the GtO and TFGO powders (~10 μm); τ is the characteristic diffusion time. Since the $^4$He atoms occupied the GO grains mainly along the carbon planes, the proportionality coefficient (the denominator in Eq. 2)) of diffusion close to the 2D case was taken to be equal to ~4.

The activation energy $E_a$ of $^4$He diffusion in GO was estimated by plotting the temperature dependence of the diffusion coefficients in the $Y=\ln(D) - X=1/T$ coordinates (see Fig. 11, typical example for GtO and TRGO-200). $E_a$ was found through a linear approximation of the experimental data in the $Y=\ln(D) - X=1/T$ coordinates, Eq. (3). The procedure was performed for the thermally activated portion of the curve at each reduction temperature

$$D = D_0 \exp\left(-\frac{E_a}{k_B T}\right) \quad (3),$$

where $D_0$ is the entropy factor depending on the frequency of collisions between the matrix and impurity molecules, $k_B$ is the Boltzmann constant. The obtained $E_a$ values for GtO and TRGO samples are shown in Fig.12.

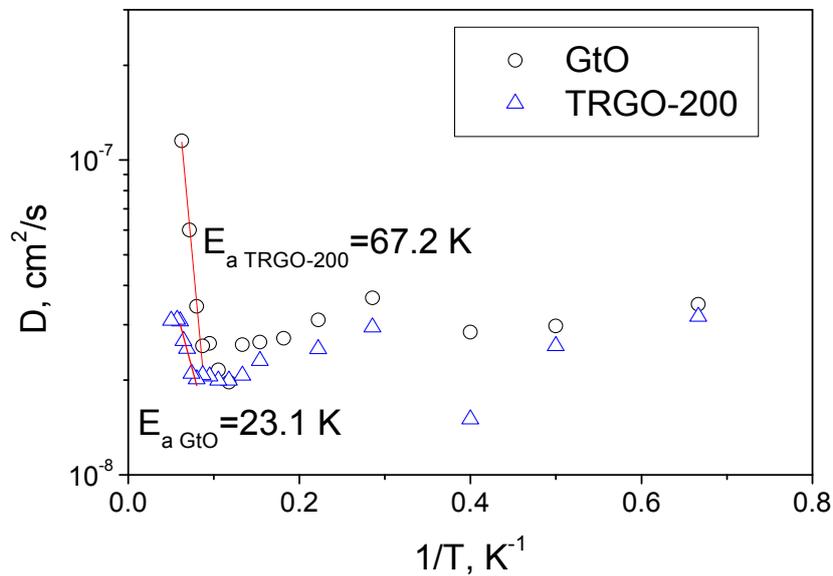

Fig. 11. The coefficients of $^4$He diffusion in GtO and TRGO-200 samples. The lines mark the linear portions of the $Y=\ln(D) - X=1/T$ dependence.

The anomalous behavior of the diffusion coefficients of $^4$He below 5 K observed for the TRGO-200 sample (such behavior has typical for other thermally reduced samples) may be caused by a transition of the impurity ($^4$He) atoms to the state of a two-dimensional quantum liquid [38].

According to XRD (see this study, section 2) and Raman spectroscopy [26] data, heating to $T=200^0$ C causes intensive evaporation of the water intercalated in the interplanar space of carbon and 'exfoliation' of the GO sheets into individual flakes [15]. The number of the interlayer cavities decreases and the influence of the other cavity wall on the $^4$He atoms diminishes significantly. As a result, the activation energy of $^4$He diffusion in the TRGO-200 sample drops sharply as against the starting GtO (Fig. 12). Heating to higher temperatures triggers several processes influencing the kinetics of helium sorption in the thermally reduced samples. Firstly, with oFGs removed, the neighboring graphene flakes 'stick' together again under the Van der Waals force [15, 39]. Besides, heating encourages relaxation of mechanical stresses and smoothes down folds and ripples [26]. These processes are favourable to the reduction of the starting layered structure of graphene oxide and raise the activation energy (see Fig. 12, $T_{reduction}$ = 300 and 500º C). On the other hand, the removal of gFGs allows the carbon atoms depart from the planes and form defects, which opens additional surface areas and ways for sorption and lowers the activation energy (Fig. 12, TRGO-700). The reduction of the layered structure and graphitization were the dominant processes in the TRGO-900 sample. The heating of GtO samples above 200º C activates two competing processes controlling the temperature dependences of the diffusion coefficients of helium, namely, the reduction of the carbon structure of graphite (Fig. 7) enhancing the activation energy and the formation of defects suppressing the activation energy. As a result, the dependence of the activation energy of helium diffusion on the temperature of the thermal reduction of graphene oxide exhibits a nonmonotonic behavior (Fig. 12), which agrees well with the XRD data (see Figs. 6 and 7).

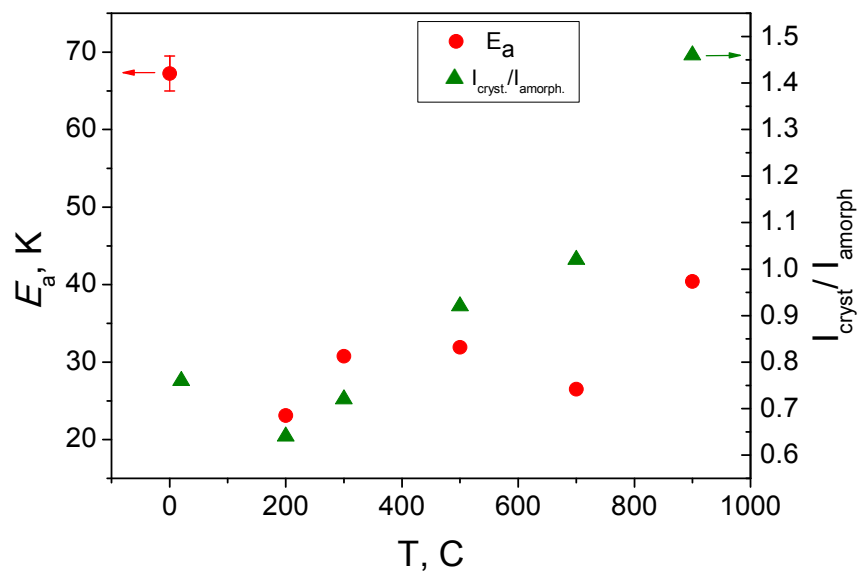

Fig. 12. The dependence of the activation energy of $^4$He diffusion (●) and growth of the crystalline TRGO phase (■) upon the temperature of thermal reduction of graphene oxide.

**Conclusions**

The effect of the reduction temperature on the kinetics of low temperature $^4$He sorption and the structural characteristics of graphene oxid has been investigated in the temperature interval 1.5 – 20 K. The time dependences of pressure variations on sorption/desorption of $^4$He are well described by the exponential one-parameter function. The times of $^4$He sorption increased for all the samples as the temperature lowered from 20 K to 11-12 K, which is typical for thermally activated diffusion. Below 5 K the characteristic times of $^4$He sorption by the GtO and TRGO-200 samples were only slightly dependent on temperature, which is indicative of the dominance of tunnel diffusion over the thermally activated mechanism. The characteristic times of the graphene oxide samples reduced at higher temperatures (TRGO-300, TRGO-500, TRGO-700 and TRGO-900) grew at T< 5 K with lowering temperature presumably because the removal of oxygen functional groups (oFGs) produced additional defects in the carbon planes and hence increased the number of diffusion-impeding potential barriers in the imperfect areas. The anomalous behavior of the diffusion coefficients of $^4$He below 5 K observed for the TRGO-300, TRGO-500, TRGO-700 and TRGO-900 samples can be caused by transformation of the impurity $^4$He atoms into the state of a two-dimensional quantum liquid.

The activation energies $Ea$ of $^4$He diffusion have been estimated for the starting and thermally treated GtO samples. In the TRGO-200 the activation energy of $^4$He diffusion decreases 2.9 times as against the starting graphite oxide due to the exfoliation of the GtO sheets caused by evaporation of the water intercalated in the interlayer space. The thermal treatment of GtO samples above 200$^o$ C triggers two competing processes accounting for the nonmonotonic behavior of the activation energy as a function of the thermal treatment temperature, namely, the recovery of the carbon structure enhancing the activation energy and the formation of defects suppressing the activation energy through opening additional surface areas and ways for sorption. The dependence of the activation energy on the thermal treatment temperature correlates well with the amount of the crystalline GtO phase varying with increasing annealing temperature.

**Acknowledgment**

Financial support from Spanish Ministry MINECO and the European Regional Development Fund (project ENE2013-48816-C5-5-R), the Regional Government of Aragon and the European Social Fund DGA-ESF (project T66) and Targeted Comprehensive Fundamental Research Program of NASU (project 6/16-H) is gratefully acknowledged.

**References**

1. D.D.L. Chung, *J. Mater.Sci.* **37**, 1475 (2002) doi: 10.1023/A:1014915307738.
2. A.K. Geim and K.S. Novoselov, *Nat Mater*. **6**(3), 183 (2007) doi:10.1038/nmat1849
3. K.I. Bolotin, K.J. Sikes, Z. Jiang, M. Klima, G. Fudenberg, J. Hone, P. Kim, H.L. Stormer, *Solid State Commun.* **146**, 351 (2008) doi:10.1016/j.ssc.2008.02.024
4. S. V. Morozov, K. S. Novoselov, M. I. Katsnelson, F. Schedin, D. C. Elias, J. A. Jaszczak, A. K. Geim, *Phys. Rev. Lett.* **100**, 016602, (2008) doi: http://dx.doi.org/10.1103/PhysRevLett.100.016602
5. C. Lee, X. D. Wei, J. W. Kysar, J. Hone, *Science* **321**, 385, (2008) doi: 10.1126/science.1157996.
6. A.A. Balandin, S. Ghosh, W. Bao, I. Calizo, D. Teweldebrhan, F. Miao, C.N. Lau, *Nano Lett.* **8**, 902, (2008) doi: 10.1021/nl0731872.
7. K.S. Novoselov, Z. Jiang, Y. Zhang, S.V. Morozov, H.L. Stormer, U. Zeitler, J. C. Maan, G. S. Boebinger, P. Kim, A. K. Geim, *Science* **315**, 1379 (2007) doi: 10.1126/science.1137201.
8. K.S. Novoselov, A.K. Geim, S.V. Morozov, D. Jiang, Y. Zhang, S.V. Dubonoc, I.V. Grigorieva, A. Firsov, *Science* **306**, 666 (2004) doi: 10.1126/science.1102896
9. K.S. Kim, Y. Zhao, H.Jang, S.Y. Lee, J.M. Kim, J.H. Ahn, P. Kim, J.Y. Choi, B.H. Hong, *Nature* **457**, 706 (2009) doi:10.1038/nature07719
10. N. Liu, F. Luo, H.X. Wu, Y.H. Liu, C. Zhang, J. Chen, *Adv. Funct. Mater.* **18**, 1518 (2008) doi: 10.1002/adfm.200700797
11. W. S. Hummers, R. E. Offermann, *J. Am. Chem. Soc.* **80**, 1339 (1958) doi: 10.1021/ja01539a017
12. L. Shahriary, A.A. Athawale, *International Jornal of Renewable Energy and Environmental Engineering* **2**, 356 (2014) doi: http://dx.doi.org/10.1155/2014/276143
13. A. Left, H. Heyong, M. Forster, J. Klimowski, *J. Phys.Chem.B* **102**, 4477 (1998) doi: 10.1021/jp9731821
14. H.Heyong, J. Klimowski, M. Forster, *Chem. Phys. Lett.* **287**, 53 (1998) http://dx.doi.org/10.1016/S0009-2614(98)00144-4
15. S.H. Huh. *Thermal Reduction of Graphene Oxide, Physics and Applications of Graphene - Experiments*, InTech, CC BY-NC-SA (2011) doi: 10.5772/14156
16. A. Ganguly, S. Sharma, P. Papakonstantinou, and J. Hamilton, *J. Phys. Chem. C* **115 (34)**, 17009 (2011) doi: 10.1021/jp203741y

17. M. D. Stoller, S. J. Park, Y. Zhu, J. An, R. S. Ruoff, *Nano Lett.* **8**, 3498 (2008) doi:10.1021/nl802558y
18. X. Wang, L. Zhi, and K. Mullen, *Nano Lett.* **8**, 323 (2008) doi: 10.1021/nl072838r
19. Q. Liu, Z. Liu, X. Zhang, L.Yang, N. Zhang, G. Pzn, S. Yin, Y.Chen, J. Wei, *Adv. Funct. Mater.* **19**, 894 (2009) doi: 10.1002/adfm.200800954
20. A.V. Dolbin, V.B. Esel'son, V.G. Gavrilko, V.G. Manzhelii, N.A. Vinnikov, R.M. Basnukaeva, V.V. Danchuk, and N.S. Mysko, E.V. Bulakh, W.K. Maser and A.M. Benito, *Fiz. Nizk. Temp.* **39**, 1397 (2013) [*Low Temp. Phys.* **39**, 1090 (2013)] http://dx.doi.org/10.1063/1.4830421
21. B. A. Danilchenko, I. I. Yaskovets, I. Y. Uvarova, A. V. Dolbin, V. B. Esel'son, R. M. Basnukaeva and N. A. Vinnikov, Appl. Phys. Lett. **104**, 173109 (2014). http://dx.doi.org/10.1063/1.4874880
22. A.V. Dolbin, V.B. Esel'son, V.G. Gavrilko, V.G. Manzhelii, N.A. Vinnikov, R.M. Basnukaeva, I.I. Yaskovets, I.Yu. Uvarova, and B.A. Danilchenko, *Fiz. Nizk. Temp.* **40**, 317 (2014) [*Low Temp. Phys.* **40**, 246 (2014)] doi: http://dx.doi.org/10.1063/1.4868528
23. A.V. Dolbin, V.B. Esel'son, V.G. Gavrilko, V.G. Manzhelii, N.A. Vinnikov, S.N. Popov, *JETP Letters* **93**, 577 (2011) doi: 10.1063/1.4830421
24. A.V. Dolbin, V.B. Esel`son, V.G. Gavrilko,V.G. Manzhelii, N.A.Vinnikov, S.N. Popov. *Fiz. Nizk. Temp.* **38**, 1216 (2012) [*Low Temp. Phys.* **38**, 962 (2012)]. doi: http://dx.doi.org/10.1063/1.4758785
25. A.V. Dolbin, V.B. Esel`son, V.G. Gavrilko,V.G. Manzhelii, N.A.Vinnikov, R. M. Basnukaeva, *Fiz. Nizk. Temp.* **39**, 475 (2013) [*Low Temp. Phys.* **39**, 370 (2013)]. http://dx.doi.org/10.1063/1.4802502
26. A.V. Dolbin, M.V. Khlistyuck, V.B. Esel'son, V.G. Gavrilko, N.A. Vinnikov, R.M. Basnukaeva, I. Maluenda, W.K. Maser and A.M. Benito, *Applied Surface Science*, **361**, 213 (2016) doi: http://dx.doi.org/10.1016/j.apsusc.2015.11.167
27. Gao, L. B. Alemany, L. Ci, P. M Ajayan, *Nature Chemistry* **1**, 403 (2009) doi: 10.1038/NCHEM.281
28. H. K. Jeong, Y. P. Lee, M. H. Jin, E. S.Kim, J. J. Bae, Y.H. Lee, *Chem. Phys. Lett* **470**, 255 (2009) doi:10.1016/j.cplett.2009.01.050
29. B. C. Brodie, *Phil. Trans. R. Soc. A.* **149**, 249 (1859) doi: 10.1098/rstl.1859.0013
30. M. Hirata, T. Gotou, S. Horiuchi, M. Fujiwara, M. Ohba, *Carbon* **42**, 2929 (2004) doi: 10.1016/j.carbon.2004.07.003
31. G. I. Titelman, V. Gelman, S. Bron, R. L. Khalfin, Y. Cohen, H.Bianco-Peled, *Carbon* **43**, 641 (2005) doi:10.1016/j.carbon.2004.10.035

32. H. M. Ju, S. H. Huh, S. H. Choi, H. L. Lee, *Mater. Lett.* **64**, 357 (2010) doi:10.1016/j.matlet.2009.11.016
33. S. Mao, H. Pu and J. Chen, *RSC Advances* **2**, 2643 (2012) doi: 10.1039/C2RA00663D
34. A.V. Dolbin, M.V. Khlistyuck, V.B. Esel'son, V.G. Gavrilko, N.A.Vinnikov, and R.M. Basnukaeva, I. Maluenda, W.K. Maser, and A.M. Benito, *Fiz. Niz. Temp.* **42**, 75 (2016) http://dx.doi.org/10.1063/1.4939155
35. C. Valles, J. David Nunez, A.M. Benito, W.K. Maser, *Carbon* **50**, 835 (2012) http://dx.doi.org/10.1016/j.carbon.2011.09.042
36. A. V. Dolbin, V. B. Esel'son, V. G. Gavrilko, V. G. Manzhelii, N. A. Vinnikov and S. N. Popov, *Fiz. Nizk. Temp.* **36**, 1352 (2010) [*Low Temp. Phys.* **36**, 1091 (2010)] doi: http://dx.doi.org/10.1063/1.3530423
37. A. V. Dolbin, V. B. Esel'son, V. G. Gavrilko, V. G. Manzhelii, N. A. Vinnikov, S. N. Popov, N. I. Danilenko and B. Sundqvist, *Fiz. Nizk. Temp.* **35**, 613 (2009) [*Low Temp. Phys.* **35**, 484 (2009)] doi: http://dx.doi.org/10.1063/1.3151995
38. S. Nakamura, K. Matsui, T. Matsui, Hiroshi Fukuyama, *Low Temp. Phys.* **171**, 711 (2013) doi: 10.1007/s10909-012-0847-5
39. M. Acik, G. Lee, C. Mattevi, A. Pirkle, R. M. Wallace, M. Chhowalla, K. Cho, Yv. Chabal, *J. Phys. Chem. C* **115**, 19761 (2011) doi: 10.1021/jp2052618